\documentclass[
    preprint,
    prb,
    amsmath,amssymb,
    aps,
    ]{revtex4-1}

\usepackage{graphicx}
\usepackage{dcolumn}
\usepackage{siunitx}
\usepackage{float}
\usepackage{bm}
\usepackage{amsmath}

\usepackage{fancyhdr}
\usepackage{fnpos}
\usepackage[english]{babel}
\usepackage{array}
\usepackage{siunitx}
\usepackage{droidsans}
\usepackage{charter}
\usepackage[T1]{fontenc}
\usepackage{setspace}

\begin{document}

\title{Ultra-narrow spin wave metasurface for focusing application}
\author{M.~Zelent$^{1}$ } \email{mateusz.zelent@amu.edu.pl}
\author{M.~Mailyan$^{2}$ } 
\author{V. Vashistha$^{1}$} 
\author{P. Gruszecki$^{1}$}
\author{O.Y.~Gorobets$^{2}$ } 
\author{Y.I.~Gorobets$^{2,3}$ } 
\author{M.~Krawczyk$^{1}$ } 
\email{krawczyk@amu.edu.pl}

\affiliation{%
$^{1}$Faculty of Physics, Adam Mickiewicz University in Poznan, Umultowska 85, Pozna\'{n}, 61-614, Poland 
}%
\affiliation{%
$^{2}$Faculty of Physics and Mathematics, National Technical University of Ukraine "Igor Sikorsky Kyiv Polytechnic Institute", 37 Peremogy ave., 03056, Kyiv, Ukraine
}%
\affiliation{%
$^{3}$Institute of Magnetism, National Academy of Sciences of Ukraine, 36-b Vernadskogo st., 03142, Kyiv, Ukraine
}%

\begin{abstract}
In this paper, we show that the phase shift of the spin waves can be controlled by metasurface formed by an ultra-narrow non-magnetic spacer separating two thin ferromagnetic films. 
For this purpose, we exploit  the strength of the exchange coupling of RKKY type between the films which allows to tune the phase of the transmitted  spin waves in the wide range of angles [$-\pi/2$;$\pi/2$]. We combined the phase-shift dependency along the interface with the lens equation to demonstrate numerically the metalens for spin waves.

\end{abstract}
\maketitle

\section{Introduction}
Spin waves (SWs) are promising data carriers for future logic devices, information processing and communication with low-energy dissipation, relatively high and flexible frequency operation and possible miniaturization down to nanoscale~\cite{Kruglyak2010a,Magnonics2013,Barman2018,Csaba2017}. However, still many basic units, such as SW generators and detectors, SW waveguides, SW amplifiers or SW modulators have to be further developed to demonstrate their useful performance~\cite{Chumak2015}. Control of the phase of propagating SWs is expected to be one of the key element in future magnonics like it  is in microwave technology, electronics, and photonics. To fulfill the requirements of miniaturization, the expected magnonic component shall be small, ideally shorter than the  wavelength. An interesting idea to fulfill this condition comes from  photonics, where the use of  ultra-thin slabs of sub-wavelength thickness, called \textbf{metasurfaces} (MSs) have been introduced. They promise valuable advantages in controlling the waves by spatial manipulation of the wavefront characteristics. 

The MS concept is based on the phase gradient introduced by the specially arranged tiny elements having a sub-wavelength size as compared with the wavelength $\lambda$ of the incident wave~\cite{holloway2012overview}. In optics, these tiny phase-shifting elements are mostly designed from metallic materials  or dielectric resonances. They rely on subwavelength gratings, resonators, waveguides, and they introduce geometric phase differences to tune the phase of the reflected or transmitted wave. Their working principle is well described with the generalized Snell's law~\cite{yu2011light} which takes into account an inhomogeneous phase change on the two-dimensional plane of the MS.
Such structures are widely investigated for  application in steering propagation of electromagnetic~\cite{BalthasarMueller2017MetasurfacePolarization} and  acoustic waves~\cite{diaz2017perfect, Li2014ExperimentalMetasurfaces,ma2014acoustic}. Originally designed at radiowave frequencies for radar and space communications~\cite{yu2014flat}, MSs have been implemented to design many new planner devices at infrared and optical frequencies, such as an ultra-thin lens, vortex plates, a polarization converter, color filters, holograms~\cite{khorasaninejad2016metalenses,lin2014dielectric,yu2014flat,ni2012broadband,sun2012gradient,vashistha2017all,jha2015metasurface}. 

\section{Analitical model}
In this paper, we propose a new concept of the MS for SWs based on the ultra-narrow interface between two thin ferromagnetic films and demonstrate numerically a possibility of exploiting this idea for design the SW phase-shifting systems. The working principle behind the proposed magnonic MS is based on the interlayer exchange coupling between magnetic films, which can be mediated by a non-magnetic metallic spacer between the edges of the films and had essentially the same physical origin as the Ruderman-Kittel-Kasuya-Yosida (RKKY) interaction. We showed analytically and numerically that such an ultra-narrow interface, of the width up to a few atomic planes, can effectively change the phase of the reflected and transmitted SWs having  tens on nanometers long wavelength. In particular, we demonstrate the magnonic MS allowing to focus the plane  SWs in an arbitrary located focal point, i.e., we show magnonic metalens. 
It is a new approach with respect to the already demonstrated concepts of SW lenses basing on either modification of the interface curvature~\cite{papp2018lens, csaba2014spin, toedt2016design} or introduction of the magnonic refractive index gradient~\cite{whitehead2018luneburg, dzyapko2016reconfigurable, gruszecki2018mirage}. Both these methods require wide lenses and exploit the classical optic rules to focus the wave.

Let's first consider the system of two semi-infinite ferromagnetic thin films separated by an ultra-narrow interface along the $y$-axis with the uniform static external magnetic field $\mathbf{H}_{0}$ which is parallel to the interface, and saturates the system. We assume that the magnetizations in both ferromagnets are exchange coupled, with the strength of the coupling described by the parameter $A_{12}$. Different models of the oscillatory interlayer exchange coupling between ferromagnetic layers separated by a non-magnetic metal have a common physical mechanism: due to contact of the spacer with the first ferromagnetic layer, the spacer conduction electrons experience spin polarization~\cite{Bruno1993InterlayerPicture,Bruno1995QuantumCo/Cu/Co001,Bruno1991OscillatorySpacer}. The latter extends throughout the spacer and interacts with the second ferromagnet and results in the effective exchange interaction. 
Depending on the assumptions, various models can be explored (RKKY, tight-binding, hole-confinement, free-electron, s-d mixing etc.)~\cite{Bruno1993InterlayerPicture,Ruderman1954IndirectElectrons,Chappert1991Long-periodPicture,Bruno1991OscillatorySpacer}.
According to the model developed in Ref. \citealp{Bruno1992Ruderman-KittelCoupling} the RKKY coupling energy per unit area with zero angles between the magnetizations in the ferromagnets is proportional to the interlayer coupling constant $J_{12}$:
\begin{equation}\label{eq:ex_const}
A_{12}=J_{12}/(M_{01}M_{02}),
\end{equation}
where $M_{01}$ and $M_{02}$ are the saturation magnetizations of the first and the second material, respectively. To relate variation of $A_{12}$ with respect to the interface width $D$ we use the model developed in Ref.~[\citealp{Bruno1995QuantumCo/Cu/Co001}].
For the interlayer exchange coupling between the two semi-infinite fcc Co layers with Cu(001) spacer, Bruno found a good agreement between the theoretical model and the experimental observations~\cite{Bruno1991OscillatorySpacer,Johnson1992StructuralLayers}. Accordingly, the interlayer coupling constant dependence on the spacer thickness $D$ [in atomic layers (AL) units] can be modeled with a simplified formula~\cite{Bruno1995QuantumCo/Cu/Co001}:
\begin{equation}\label{eq:ex_integral}
J_{12}=\frac{I_1}{D^2}\sin\left(\frac{2\pi}{\Lambda_1}+\phi_1\right)+\frac{I_2}{D^2}\sin\left(\frac{2\pi}{\Lambda_2}+\phi_2\right),
\end{equation}
where $I_1$ and $I_2$ are the coefficients describing the coupling strength between ferromagnets, $\Lambda_1$ and $\Lambda_2$ are the oscillation periods, $\phi_1$ and $\phi_2$ are the phases of the oscillatory coupling. Eq.~(\ref{eq:ex_integral}) reproduces well the experimental data of Co/Cu(001)/Co multilayers with $\Lambda_1=2.6$ AL and $\Lambda_2=8.0$ AL~\cite{Johnson1992StructuralLayers,Bruno1995QuantumCo/Cu/Co001}; coupling strengths $I_1=17.7$ erg/cm$^2$ and $I_2=4.3$ erg/cm$^2$~\cite{Bloemen1993ShortCo/Cu/Co100}; phases $\phi_1=\pi/2$ and $\phi_2=\pi$~\cite{Bruno1995QuantumCo/Cu/Co001} (see, Fig.~\ref{Phase_on_aex_and_d} b).

To calculate the phase shift $\varphi_T$ and the transmission coefficient $T$ of the SW transmitted through the interface connecting the two ferromagnetic thin films we use an analytical approach for the SW propagation in the system of two ferromagnetic films separated by an infinitely thin non-magnetic interface.
Based on the Landau-Lifshitz equations [Eq.~(S1)  in Supplemental Materials] in linear approximation, assuming the plane wave solutions with wavenumbers $k_1$ and $k_2$ homogeneous across the film thickness [Eq.~(S4)] we solve the system of equations obtained from the exchange boundary conditions at the interface $x =0$~\cite{Kruglyak2014MagnetizationThickness,gruszecki2017goos}:
\begin{equation}\label{eq:boundary_conditions}
\begin{cases}
\left(A_{12}(m_{2n}-\xi{m}_{1n})+\alpha_1\cfrac{\partial m_{1n}}{\partial x}\right)\bigg\vert_{x=0}=0, \\
\left(A_{12}(m_{1n}-\cfrac{{m}_{2n}}{\xi})-\alpha_2\cfrac{\partial m_{2n}}{\partial x}\right)\bigg\vert_{x=0}=0,
\end{cases}
\end{equation}
where $\xi=M_{02}/M_{01}$, $n=x,z$. $m_{jn}$ is a dynamic magnetization component and $\alpha_j$ is the exchange parameter in the $j$-th ferromagnet, $j =1, 2$: $\alpha_j=A_{{\text{ex}_{j}}}/M_{0j}^2$, where $A_{{\text{ex}_{j}}}$ is the $j$-th ferromagnet exchange stiffness constant.
We end with the following formulas:
\begin{equation}\label{eq:shift_intensity}
\begin{split}
&\varphi_T=\text{arctan}\left(A_{12}\frac{\xi\alpha_2k_2+\alpha_1k_1/\xi}{\alpha_1\alpha_2k_1k_2}\right)-
\begin{cases}
\pi, \text{ for }A_{12}<0\\0,  \text{ for } A_{12}>0
\end{cases}
\\&T=\frac{4\alpha_1^2k_1^2}{(\alpha_1\alpha_2k_1k_2/A_{12})^2+(\xi\alpha_2k_2+\alpha_1k_1/\xi)^2}.
\end{split}
\end{equation}
 Based on Eqs.~(\ref{eq:shift_intensity}) we will obtain the phase shift and intensity of the transmitted SW in dependence on $A_{12}$, which through Eq.~(\ref{eq:ex_integral}) depends also on $D$. The assumption of the infinitely narrow interface is justified whenever $\lambda \ll D$, as we will see the condition fulfilled to the large extent in the presented results. The analytical results from Eq.~(\ref{eq:shift_intensity}) will be directly compared to micromagnetic simulations results. 

 \begin{figure}[h]
 \centering
     \includegraphics[width=\columnwidth]{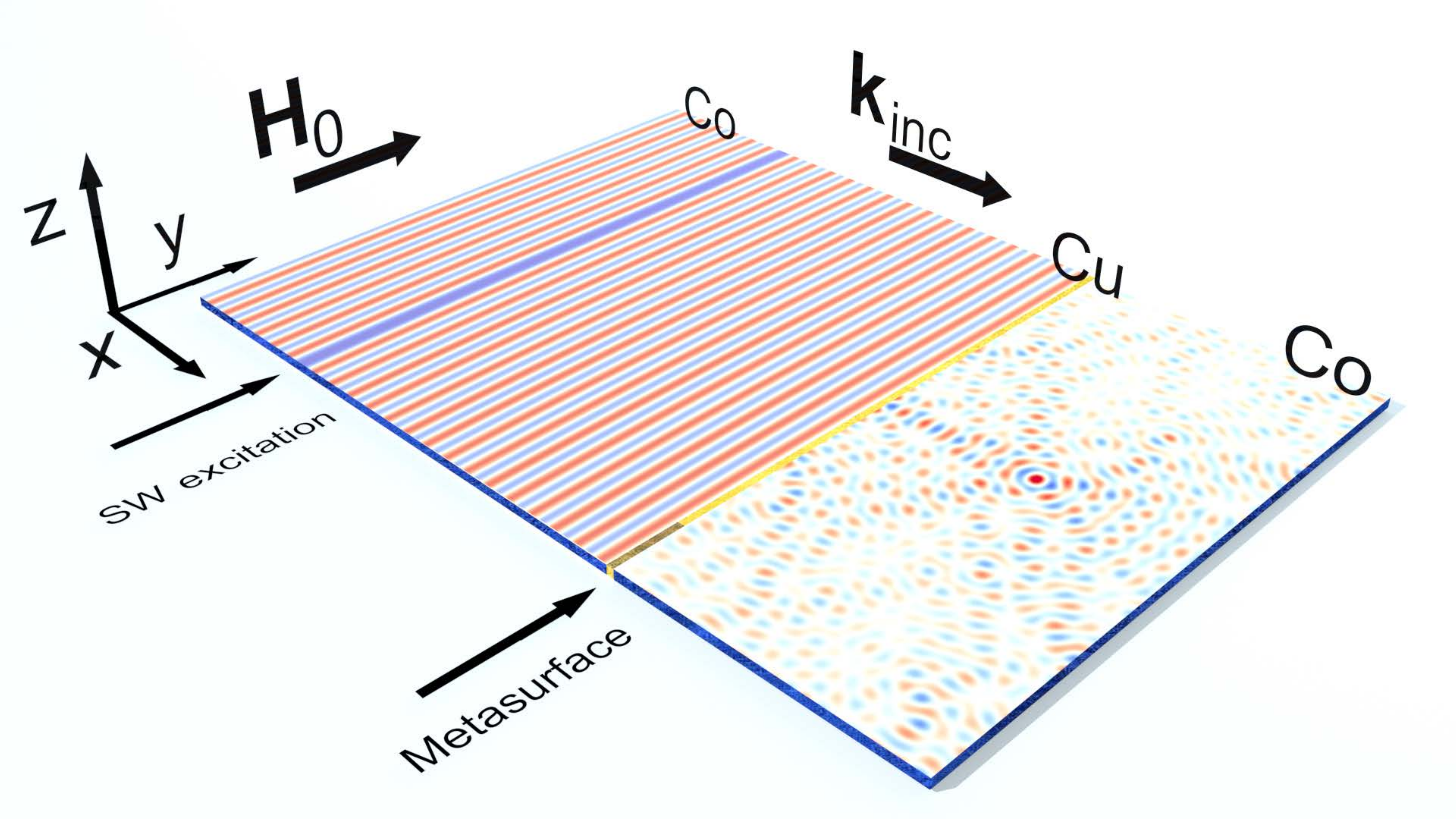} 
     \caption{ Schematic of the system design with a focusing metasurface. The system involves the two ferromagnetic thin films of Co with a non-magnetic metallic material (Cu) of varying width  along the interface (metasurface).}\label{geometry}
 \end{figure}
\section{Micromagnetic simulations}
To demonstrate the MS for SWs 
we perform the finite-difference time-domain micromagnetic simulations with Mumax$^3$ solver~\cite{MuMax2011_main,mumax_2014,Mulkers2017,Exl2014,Leliaert2014}, taking into account exchange and dipolar interactions with damping neglected~\cite{mumax_2014}. We assume homogeneous Co film of 10 nm thickness, 800 nm width and 3 $\mu$m long (along the $x, y$ and $z$, respectively), magnetized to saturation by the external magnetic field $H_0=1$ kOe oriented along the $y$ axis. We excite the harmonic SW plane waves at 80 GHz frequency (which relate to the 28 nm of the wavelength) on the left side of the interface (1 $\mu$m from the interface, see Fig.~\ref{geometry}) and record the data from the whole structure in the following time steps until the waves reach the right side of the simulation area where the absorbing boundary has been implemented~\cite{venkat2018absorbing}. The interface has been implemented as the unit cell of the discretized mesh with the artificially introduced exchange interaction $A_{12\text{ex}}$ scaled suitably to use the $A_{12}$ parameter from the analytical model from the Eq.~(\ref{eq:shift_intensity})~\footnote{We used a uniformly discretizing grid with the cell size 0.5-1.0 nm $\times$ 0.5-1.0 nm $\times$ 10 nm along the $x$, $y$ and $z$ axis, respectively.}:
$ A_{12\text{ex}}= \Delta A_{12} M_{0}^2 / 2A_{\text{ex}}$, where $\Delta$ is the lateral size of the unit cell in micromagnetic simulations. For the simulations the following parameters for Co have been taken: $M_{01}=M_{02}=M_{0}=1422$ kA/m, $A_{\text{ex},1}=A_{\text{ex},2}=A_{\text{ex}}=3\times10^{-11}$ J/m~\cite{Chen2010MicrowaveAnisotropy}, uniaxial magnetocrystalline anisotropy constant $K_1=K_2=K=4.5\times10^6$ erg/cm$^3$~\cite{Cullity2008IntroductionMaterials}, and $\Delta=0.75$ nm. 

\begin{figure}[h]
    \includegraphics[width=\columnwidth]{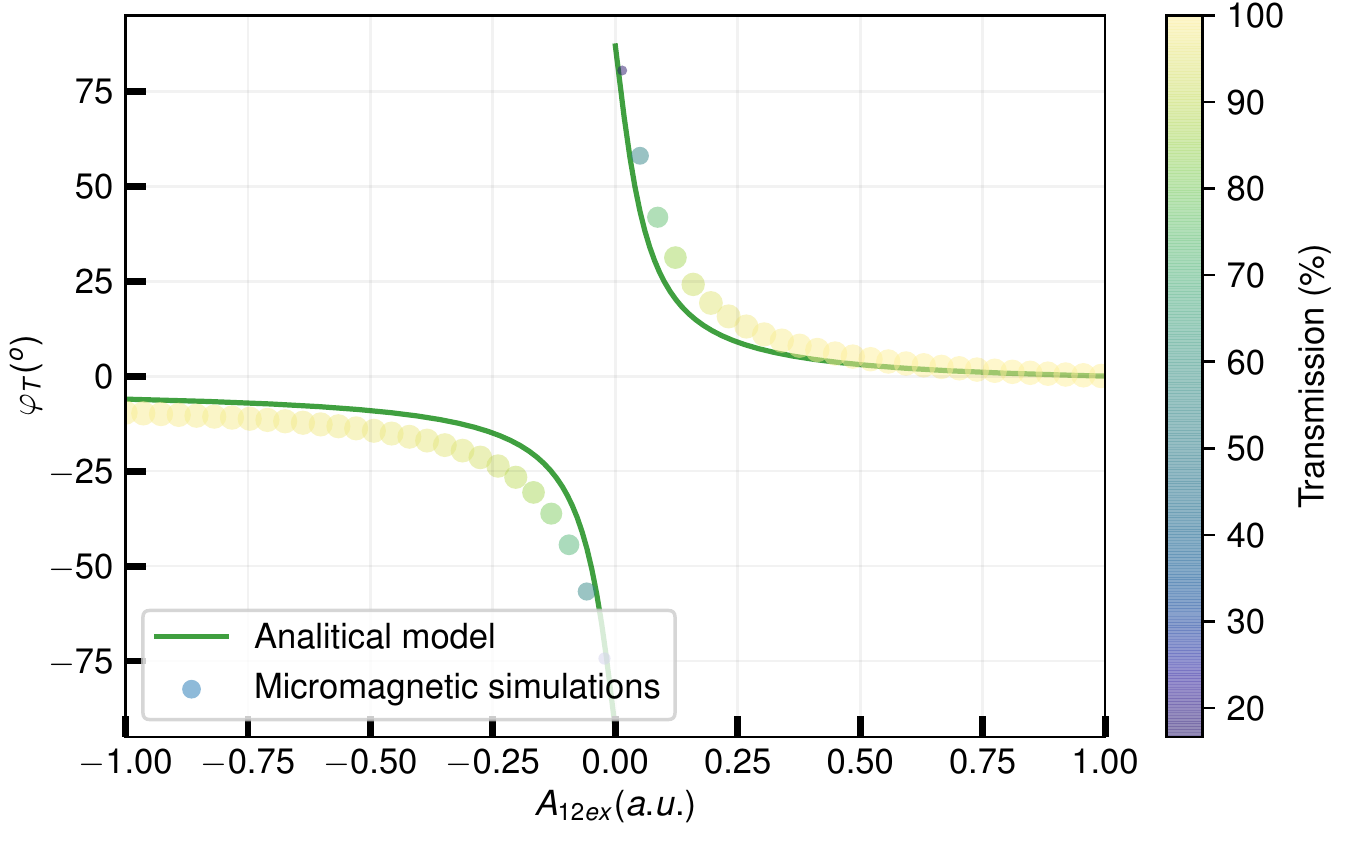}
    \caption{
    The color scaled dots represent the simulated dependence of the phase shift between the transmitted and incident SWs on the $A_{12}$ exchange parameter for 80 GHz waves. Color and size of the dots corresponds to the intensity of the transmitted SWs. The solid green line indicates the results of the analytical model.}
    \label{Phase_on_aex}
\end{figure} 
In Fig.~\ref{Phase_on_aex} we show the phase-shift and transmission of the SW transmitted through the interface in dependence on the interlayer exchange coupling constant $A_{12}$ obtained from simulations and from Eq.~(\ref{eq:shift_intensity}). Micromagnetic simulation results (dots) are in very good agreement with the analytical model (green line).
The positive and negative values of $A_{12}$ relate to ferromagnetic and antiferromagnetic coupling between the Co films, respectively. We see, that the  $A_{12}$ determines the value of the phase shift, and also the intensity of the SW transmission (superimposed in the figure with the color map). $T$ decreases with $|A_{12}|$ approaching 0, i.e., a limit which indicates exchange decoupled materials. Interestingly, at small $A_{12}$ the most significant changes of the $\varphi_T$ with $A_{12}$ are observed, indicating the region suitable for further exploitation.

\begin{figure}[h]
\centering
\includegraphics[width=\columnwidth]{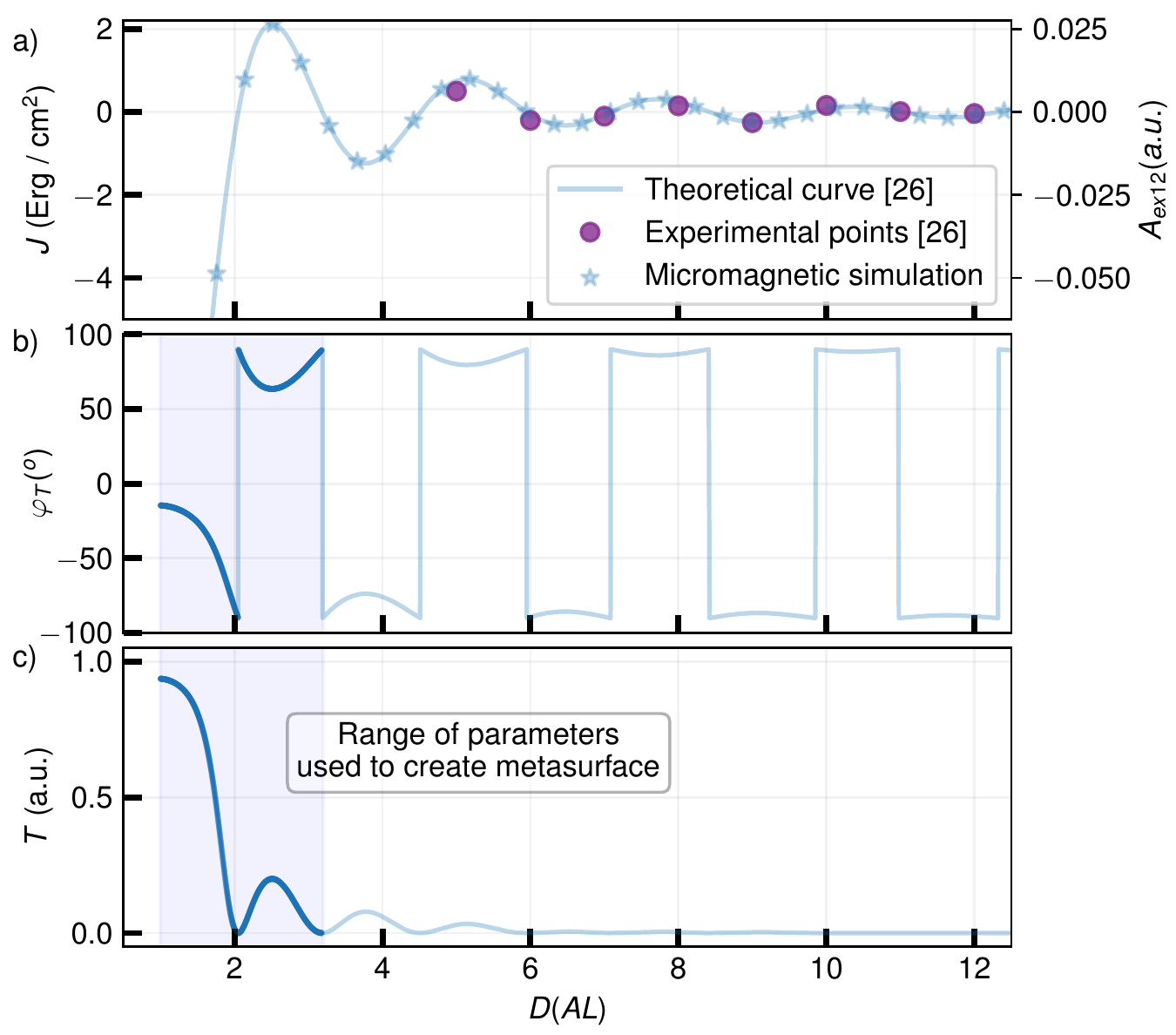}
\caption{
 Dependence of the selected parameters on the number of atomic layers on the interface: (a) the strength of the interlayer coupling constant $J$ (the exchange coupling parameter $A_{12\text{ex}}$ used in micromagnetic simulations), (b) spin wave phase shift and (c) transmittance as a function of $D$. In (a) the $J(D)$ dependence is based on the theoretical estimation for RKKY-coupling parameter proposed by P. Bruno in Ref.~[\citealt{Bruno1995QuantumCo/Cu/Co001}], the bright dots correspond to the micromagnetic simulation results, while the violet dots correspond to the experimentally measured values~\cite{Bruno1995QuantumCo/Cu/Co001}. Bold lines inside the highlighted region in the panels (b) and (c) indicate the values used to design lensing metasurface. 
}\label{Phase_on_aex_and_d}
\end{figure} 

According to Eq.~(\ref{eq:ex_integral}) the interlayer exchange constant depends on $D$. This dependence for Co/Cu/Co multilayer is shown in Fig.~\ref{Phase_on_aex_and_d} (a), where very good agreement with experimental results and perfect with micromagnetic simulations are demonstrated.
The SW phase shift and transmittance related to the dependence of $A_{12}$ on $D$ are shown in Fig.~\ref{Phase_on_aex_and_d} (b) and (c), respectively.
The largest change of the phase shift with still valuable transmission  are observed in the widths of the spacer up to 2 AL. These results clearly show that the spacers of the few atomic layers thick can determine the phase acquired by the transmitted SW in the range from -90$^\circ$ up to 90$^\circ$~\footnote{The recent investigations Ref.\cite{Szalowski2008PRB_RKKY,Hjorvarsson1997PRL_Reversible,Klose1997PRL_Continous,Unguris1997PRL_Determination} show that, the smooth changes of the RKKY interactions are feasible.}. This range of $A_{12}(D)$ will be further exploited to design lens for SWs.

The phase gradient introduced along the line of the MS will provide an effective wave vector along the interface that is imparted to the transmitted waves. This effect is described by the generalized Snell law~\cite{yu2011light}:
\begin{equation}
\sin(\theta_t)n_t - \sin(\theta_i)n_i = \frac{\lambda}{2\pi}\frac{\text{d}\varphi_T}{\text{d}y},
\end{equation}
where $\theta_i$ and $\theta_t$ are angles of incident and refracted waves with respect to the normal to the interface, and $n_i$, $ n_t$ are refractive indexes of the materials. 
This additional momentum defined in the phase gradient term $\dfrac{\text{d}\varphi_T}{\text{d}y}$ is realized in photonic MS by nanoantennas placed on the surface. 
The Snell’s law describes also the refraction of SWs~\cite{stigloher2016snell}, therefore, the phase discontinuity approach used in photonics can be expected to give similar effects for SWs. We propose to use the exchange coupling introduced above to provide a phase discontinuity for SWs. 

The generalized Snell law indicates, that the transmitted  waves can be bent into the arbitrary direction depending on  the phase gradient profile along the MS, as well as the refractive indices of the surrounding media~\cite{Yu2014}.
Thus, the phase-shift on MS can be used to design interfaces offering different functionality, including \textbf{focusing}.

The flat interface will work as a metalens, if for a given focal length $f$, the phase shift $\varphi_{T}(y)$ imposed on every point of the interface along the $y$ axis will satisfy the following equation:
\begin{equation} \label{eq:flat_lenst_eq}
\varphi_{T}(y)=-\frac{2\pi}{ \lambda} (\sqrt[]{(y-y_{0})^2+f^2}-f).
\end{equation}
 To impose this flat lens phase profile $\varphi_{T}(y)$  we assume, that the width of the spacer $D$ will be dependent on $y$. In micromagnetic simulations instead of $D(y)$ we used a discrete change of the exchange scale parameter [$A_{12}(y)$, with the step of 1 nm]\footnote{We have used the modified code of the mumax3 allowing us to use more regions to implement the scale exchange interface. The limit in the standard version of mumax3 is 256 regions.}, which is directly related to $D(y)$ according to Eqs.~(\ref{eq:ex_const}) and (\ref{eq:ex_integral}).

\begin{figure}[ht]
    \centering
    \includegraphics[width=\columnwidth]{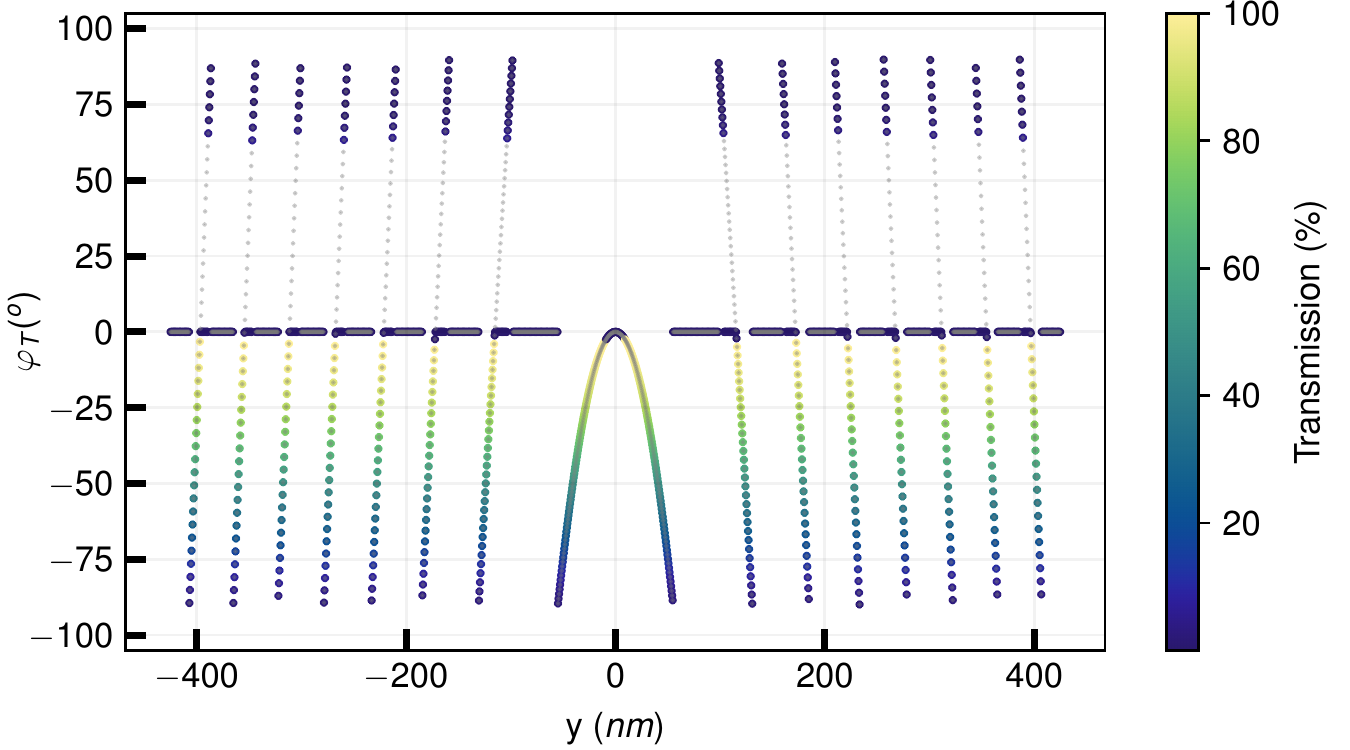} 
    \caption{The phase-shift profile of the meta-lens for SW of the wavelength $\lambda =\SI{28}{\nano\metre}$ ~and for the focal point $f=\SI{200}{\nano\metre}$. The transmission of SW is indicated by the dots color and size.}
    \label{Phase_profile}
\end{figure}

To design meta-lens based on the interlayer coupling at Co/Cu/Co interface shown in Fig.~\ref{Phase_on_aex_and_d}(a), we select its width in the range up to 3 ALs. This allows to shape the phase of the transmitted wave in a relatively wide range with the still valuable transmission, see, Fig.~\ref{Phase_on_aex_and_d}(b) and (c), although, the angles of the phase shift from 0 to 65$^\circ$ are not accessible. To increase the intensity of the focused SWs we use 14 repetitions of the available angles along the interface, distributed symmetrically on both sides of the MS center. The designed profile of the $\varphi(y)$ to focus the wave along the bisector line of the MS at 150 nm distance from MS is shown in Fig.~\ref{Phase_profile}.
The spatial dependence of the transmission of SWs through the MS is indicated by the color map. Interestingly, the whole transmittance is about $40\%$ and this low value is related to the large area of the zero-transmission regions marked with the dark horizontal lines [at $\varphi_T(y) = 0$[].

\begin{figure}[!ht]
\includegraphics[width=\columnwidth]{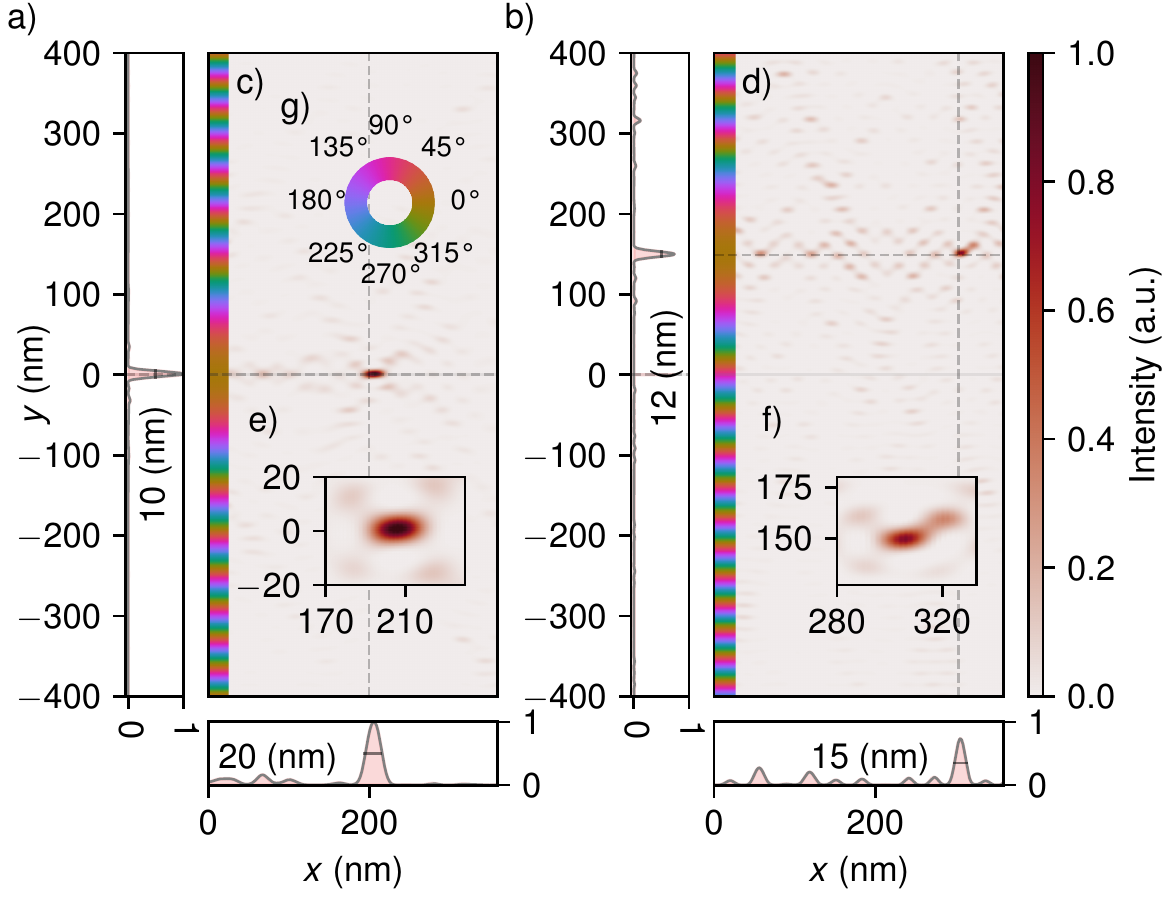} 
\caption{
The intensity of SWs (normalized to the maximum energy in the focal plane) of the transmitted waves on the $x-y$ plane. (a) The meta-lens for SWs with the focal spot centered at $y = 0$ and $200$ nm behind the MS along the $x$ axis. The matasurface is based on the phase shift design shown in Fig.~\ref{Phase_profile}.  (b) The meta-lens with the focal spot shifted to $x = 300$ and $y= 150$ nm. 
c) and d) The phase change along the interface  obtained from the lens equation to create meta-lens. The total width of the waveguide with the focusing MS is 800 nm.
}
\label{lens}
\end{figure} 


The numerical demonstration of our focusing MS with  designed $A_{12}(y)$to obtain intense small focal spot of SWs at $y=0$, $x=200$ nm (according to the phase profile shown in Fig.~\ref{Phase_profile}) is shown in Fig.~\ref{lens} (a). This result clearly shows that the SW focusing is possible and the intensity at the focal spot significantly stands out of the SW landscape.
In the focal point the SW intensity  features 7 times raised amplitude in comparison to the whole interference pattern on the right side of the interface. The half-width at the full maximum of the spot size is 20 and 10 nm along the $x$ and $y$ axis, respectively.

The phase profile $\varphi(y)$ and the respective $A_{12}(y)$ change along the $y$ axis can be designed to focus the wave also at an arbitrary position on the film. In Fig.~\ref{lens} (b) we show the micromagnetic simulation results demonstrating the off-line focusing of SWs. The intensity at focal point (located at $y=150$, $x=300$ nm) decreases to 75\% of the intensity in (a), but the focal size has similar  size (15, 12 nm along the $x$ and $y$ axis, respectively).

The efficiency, defined as the ratio of intensity in the focal region (circular area with radius of about two times of full width at half maximum~\cite{Khorasaninejad2016Polarization-InsensitiveWavelengths}) to incident power is about 20$\%$ in the case under investigation. The efficiency of the SWs concentration is dependent on the design, available angles of the phase shift, and is also limited by the loss of transmission through the interface. 
Those parameters can be further optimized and proposed metalens used for various applications in magnonics. 

\section{Conclusion}
In conclusion, we have shown in micromagnetic simulations the focusing of the  SWs in thin Co film by ultra-narrow flat phase-shifting metasurface. The proposed metasurface exploits the RKKY exchange coupling through the Cu spacer between Co thin ferromagnetic films to achieve the required phase shift profile along the interface. We used the RKKY exchange coupling estimated from the experimental data  and interpolated to a continuous variation on the spacer width. With properly designed metasurface the substantial focusing of the SWs can be achieved at the arbitrarily selected point on the film. The effectiveness of the focusing depends on the correct representation of the phase profile along the interface and the type of the exchange coupling. Based on the analytical model we conclude, that the other mechanisms of the magnetization coupling between ferromagnets can be also exploited for designing the metasurfaces for SWs.

\section{Acknowledgments}
The project is financed by the European Union Horizon 2020 Research and Innovation Program under Marie Sklodowska-Curie grant agreement No.~644348. The financial assistance from the National Science Center of Poland (MagnoWa UMO-2012/07/E/ST3/00538) and from the Polish Ministry of Science and Higher Education resources for science in 2017-2019 (project W28/H2020/2017) are also acknowledged. M. Z. also acknowledges support from Adam Mickiewicz University Foundation. The simulations were partially performed at the Poznan Supercomputing and Networking Center (Grant No.~209).

\bibliographystyle{apsrev4-1}

\begin{thebibliography}{52}%
\makeatletter
\providecommand \@ifxundefined [1]{%
 \@ifx{#1\undefined}
}%
\providecommand \@ifnum [1]{%
 \ifnum #1\expandafter \@firstoftwo
 \else \expandafter \@secondoftwo
 \fi
}%
\providecommand \@ifx [1]{%
 \ifx #1\expandafter \@firstoftwo
 \else \expandafter \@secondoftwo
 \fi
}%
\providecommand \natexlab [1]{#1}%
\providecommand \enquote  [1]{``#1''}%
\providecommand \bibnamefont  [1]{#1}%
\providecommand \bibfnamefont [1]{#1}%
\providecommand \citenamefont [1]{#1}%
\providecommand \href@noop [0]{\@secondoftwo}%
\providecommand \href [0]{\begingroup \@sanitize@url \@href}%
\providecommand \@href[1]{\@@startlink{#1}\@@href}%
\providecommand \@@href[1]{\endgroup#1\@@endlink}%
\providecommand \@sanitize@url [0]{\catcode `\\12\catcode `\$12\catcode
  `\&12\catcode `\#12\catcode `\^12\catcode `\_12\catcode `\%12\relax}%
\providecommand \@@startlink[1]{}%
\providecommand \@@endlink[0]{}%
\providecommand \url  [0]{\begingroup\@sanitize@url \@url }%
\providecommand \@url [1]{\endgroup\@href {#1}{\urlprefix }}%
\providecommand \urlprefix  [0]{URL }%
\providecommand \Eprint [0]{\href }%
\providecommand \doibase [0]{http://dx.doi.org/}%
\providecommand \selectlanguage [0]{\@gobble}%
\providecommand \bibinfo  [0]{\@secondoftwo}%
\providecommand \bibfield  [0]{\@secondoftwo}%
\providecommand \translation [1]{[#1]}%
\providecommand \BibitemOpen [0]{}%
\providecommand \bibitemStop [0]{}%
\providecommand \bibitemNoStop [0]{.\EOS\space}%
\providecommand \EOS [0]{\spacefactor3000\relax}%
\providecommand \BibitemShut  [1]{\csname bibitem#1\endcsname}%
\let\auto@bib@innerbib\@empty
\bibitem [{\citenamefont {Kruglyak}\ \emph {et~al.}(2010)\citenamefont
  {Kruglyak}, \citenamefont {Demokritov},\ and\ \citenamefont
  {Grundler}}]{Kruglyak2010a}%
  \BibitemOpen
  \bibfield  {author} {\bibinfo {author} {\bibfnamefont {V.~V.}\ \bibnamefont
  {Kruglyak}}, \bibinfo {author} {\bibfnamefont {S.~O.}\ \bibnamefont
  {Demokritov}}, \ and\ \bibinfo {author} {\bibfnamefont {D.}~\bibnamefont
  {Grundler}},\ }\href@noop {} {\bibfield  {journal} {\bibinfo  {journal} {J.
  Phys. D: Appl. Phys.}\ }\textbf {\bibinfo {volume} {43}},\ \bibinfo {pages}
  {264001} (\bibinfo {year} {2010})}\BibitemShut {NoStop}%
\bibitem [{\citenamefont {Demokritov}\ and\ \citenamefont
  {Slavin}(2013)}]{Magnonics2013}%
  \BibitemOpen
  \bibinfo {editor} {\bibfnamefont {S.~O.}\ \bibnamefont {Demokritov}}\ and\
  \bibinfo {editor} {\bibfnamefont {A.~N.}\ \bibnamefont {Slavin}},\ eds.,\
  \href {\doibase 10.110.1007/978-3-642-30247-3} {\emph {\bibinfo {title}
  {Magnonics. From Fundamentals to Applications}}},\ \bibinfo {series} {Topics
  in Applied Physics}, Vol.\ \bibinfo {volume} {125}\ (\bibinfo  {publisher}
  {Springer},\ \bibinfo {year} {2013})\BibitemShut {NoStop}%
\bibitem [{\citenamefont {Barman}\ and\ \citenamefont
  {Sinha}(2018)}]{Barman2018}%
  \BibitemOpen
  \bibinfo {editor} {\bibfnamefont {A.}~\bibnamefont {Barman}}\ and\ \bibinfo
  {editor} {\bibfnamefont {J.}~\bibnamefont {Sinha}},\ eds.,\ \href {\doibase
  10.1007/978-3-319-66296-1} {\emph {\bibinfo {title} {Spin Dynamics and
  Damping in Ferromagnetic Thin Films and Nanostructures}}}\ (\bibinfo
  {publisher} {Springer International Publishing},\ \bibinfo {year}
  {2018})\BibitemShut {NoStop}%
\bibitem [{\citenamefont {Csaba}\ \emph {et~al.}(2017)\citenamefont {Csaba},
  \citenamefont {Papp},\ and\ \citenamefont {Porod}}]{Csaba2017}%
  \BibitemOpen
  \bibfield  {author} {\bibinfo {author} {\bibfnamefont {G.}~\bibnamefont
  {Csaba}}, \bibinfo {author} {\bibfnamefont {Ã.}~\bibnamefont {Papp}}, \ and\
  \bibinfo {author} {\bibfnamefont {W.}~\bibnamefont {Porod}},\ }\href
  {\doibase 10.1016/j.physleta.2017.02.042} {\bibfield  {journal} {\bibinfo
  {journal} {Physics Letters, Section A: General, Atomic and Solid State
  Physics}\ }\textbf {\bibinfo {volume} {381}},\ \bibinfo {pages} {1471}
  (\bibinfo {year} {2017})}\BibitemShut {NoStop}%
\bibitem [{\citenamefont {Chumak}\ \emph {et~al.}(2015)\citenamefont {Chumak},
  \citenamefont {Vasyuchka}, \citenamefont {Serga},\ and\ \citenamefont
  {Hillebrands}}]{Chumak2015}%
  \BibitemOpen
  \bibfield  {author} {\bibinfo {author} {\bibfnamefont {A.~V.}\ \bibnamefont
  {Chumak}}, \bibinfo {author} {\bibfnamefont {V.~I.}\ \bibnamefont
  {Vasyuchka}}, \bibinfo {author} {\bibfnamefont {A.~A.}\ \bibnamefont
  {Serga}}, \ and\ \bibinfo {author} {\bibfnamefont {B.}~\bibnamefont
  {Hillebrands}},\ }\href@noop {} {\bibfield  {journal} {\bibinfo  {journal}
  {Nature Physics}\ }\textbf {\bibinfo {volume} {11}},\ \bibinfo {pages} {453}
  (\bibinfo {year} {2015})}\BibitemShut {NoStop}%
\bibitem [{\citenamefont {Holloway}\ \emph {et~al.}(2012)\citenamefont
  {Holloway}, \citenamefont {Kuester}, \citenamefont {Gordon}, \citenamefont
  {O'Hara}, \citenamefont {Booth},\ and\ \citenamefont
  {Smith}}]{holloway2012overview}%
  \BibitemOpen
  \bibfield  {author} {\bibinfo {author} {\bibfnamefont {C.~L.}\ \bibnamefont
  {Holloway}}, \bibinfo {author} {\bibfnamefont {E.~F.}\ \bibnamefont
  {Kuester}}, \bibinfo {author} {\bibfnamefont {J.~A.}\ \bibnamefont {Gordon}},
  \bibinfo {author} {\bibfnamefont {J.}~\bibnamefont {O'Hara}}, \bibinfo
  {author} {\bibfnamefont {J.}~\bibnamefont {Booth}}, \ and\ \bibinfo {author}
  {\bibfnamefont {D.~R.}\ \bibnamefont {Smith}},\ }\href@noop {} {\bibfield
  {journal} {\bibinfo  {journal} {IEEE Antennas and Propagation Magazine}\
  }\textbf {\bibinfo {volume} {54}},\ \bibinfo {pages} {10} (\bibinfo {year}
  {2012})}\BibitemShut {NoStop}%
\bibitem [{\citenamefont {Yu}\ \emph {et~al.}(2011)\citenamefont {Yu},
  \citenamefont {Genevet}, \citenamefont {Kats}, \citenamefont {Aieta},
  \citenamefont {Tetienne}, \citenamefont {Capasso},\ and\ \citenamefont
  {Gaburro}}]{yu2011light}%
  \BibitemOpen
  \bibfield  {author} {\bibinfo {author} {\bibfnamefont {N.}~\bibnamefont
  {Yu}}, \bibinfo {author} {\bibfnamefont {P.}~\bibnamefont {Genevet}},
  \bibinfo {author} {\bibfnamefont {M.~A.}\ \bibnamefont {Kats}}, \bibinfo
  {author} {\bibfnamefont {F.}~\bibnamefont {Aieta}}, \bibinfo {author}
  {\bibfnamefont {J.-P.}\ \bibnamefont {Tetienne}}, \bibinfo {author}
  {\bibfnamefont {F.}~\bibnamefont {Capasso}}, \ and\ \bibinfo {author}
  {\bibfnamefont {Z.}~\bibnamefont {Gaburro}},\ }\href@noop {} {\bibfield
  {journal} {\bibinfo  {journal} {Science}\ }\textbf {\bibinfo {volume}
  {334}},\ \bibinfo {pages} {333} (\bibinfo {year} {2011})}\BibitemShut
  {NoStop}%
\bibitem [{\citenamefont {Balthasar~Mueller}\ \emph {et~al.}(2017)\citenamefont
  {Balthasar~Mueller}, \citenamefont {Rubin}, \citenamefont {Devlin},
  \citenamefont {Groever},\ and\ \citenamefont
  {Capasso}}]{BalthasarMueller2017MetasurfacePolarization}%
  \BibitemOpen
  \bibfield  {author} {\bibinfo {author} {\bibfnamefont {J.~P.}\ \bibnamefont
  {Balthasar~Mueller}}, \bibinfo {author} {\bibfnamefont {N.~A.}\ \bibnamefont
  {Rubin}}, \bibinfo {author} {\bibfnamefont {R.~C.}\ \bibnamefont {Devlin}},
  \bibinfo {author} {\bibfnamefont {B.}~\bibnamefont {Groever}}, \ and\
  \bibinfo {author} {\bibfnamefont {F.}~\bibnamefont {Capasso}},\ }\href
  {\doibase 10.1103/PhysRevLett.118.113901} {\bibfield  {journal} {\bibinfo
  {journal} {Physical Review Letters}\ }\textbf {\bibinfo {volume} {118}},\
  \bibinfo {pages} {113901} (\bibinfo {year} {2017})}\BibitemShut {NoStop}%
\bibitem [{\citenamefont {Diaz-Rubio}\ and\ \citenamefont
  {Tretyakov}(2017)}]{diaz2017perfect}%
  \BibitemOpen
  \bibfield  {author} {\bibinfo {author} {\bibfnamefont {A.}~\bibnamefont
  {Diaz-Rubio}}\ and\ \bibinfo {author} {\bibfnamefont {S.~A.}\ \bibnamefont
  {Tretyakov}},\ }\href@noop {} {\bibfield  {journal} {\bibinfo  {journal}
  {arXiv preprint arXiv:1702.05872}\ } (\bibinfo {year} {2017})}\BibitemShut
  {NoStop}%
\bibitem [{\citenamefont {Li}\ \emph {et~al.}(2014)\citenamefont {Li},
  \citenamefont {Jiang}, \citenamefont {Li}, \citenamefont {Liang},
  \citenamefont {Zou}, \citenamefont {Yin},\ and\ \citenamefont
  {Cheng}}]{Li2014ExperimentalMetasurfaces}%
  \BibitemOpen
  \bibfield  {author} {\bibinfo {author} {\bibfnamefont {Y.}~\bibnamefont
  {Li}}, \bibinfo {author} {\bibfnamefont {X.}~\bibnamefont {Jiang}}, \bibinfo
  {author} {\bibfnamefont {R.-q.}\ \bibnamefont {Li}}, \bibinfo {author}
  {\bibfnamefont {B.}~\bibnamefont {Liang}}, \bibinfo {author} {\bibfnamefont
  {X.-y.}\ \bibnamefont {Zou}}, \bibinfo {author} {\bibfnamefont {L.-l.}\
  \bibnamefont {Yin}}, \ and\ \bibinfo {author} {\bibfnamefont {J.-c.}\
  \bibnamefont {Cheng}},\ }\href {\doibase 10.1103/PhysRevApplied.2.064002}
  {\bibfield  {journal} {\bibinfo  {journal} {Physical Review Applied}\
  }\textbf {\bibinfo {volume} {2}},\ \bibinfo {pages} {064002} (\bibinfo {year}
  {2014})}\BibitemShut {NoStop}%
\bibitem [{\citenamefont {Ma}\ \emph {et~al.}(2014)\citenamefont {Ma},
  \citenamefont {Yang}, \citenamefont {Xiao}, \citenamefont {Yang},\ and\
  \citenamefont {Sheng}}]{ma2014acoustic}%
  \BibitemOpen
  \bibfield  {author} {\bibinfo {author} {\bibfnamefont {G.}~\bibnamefont
  {Ma}}, \bibinfo {author} {\bibfnamefont {M.}~\bibnamefont {Yang}}, \bibinfo
  {author} {\bibfnamefont {S.}~\bibnamefont {Xiao}}, \bibinfo {author}
  {\bibfnamefont {Z.}~\bibnamefont {Yang}}, \ and\ \bibinfo {author}
  {\bibfnamefont {P.}~\bibnamefont {Sheng}},\ }\href@noop {} {\bibfield
  {journal} {\bibinfo  {journal} {Nature materials}\ }\textbf {\bibinfo
  {volume} {13}},\ \bibinfo {pages} {873} (\bibinfo {year} {2014})}\BibitemShut
  {NoStop}%
\bibitem [{\citenamefont {Yu}\ and\ \citenamefont
  {Capasso}(2014{\natexlab{a}})}]{yu2014flat}%
  \BibitemOpen
  \bibfield  {author} {\bibinfo {author} {\bibfnamefont {N.}~\bibnamefont
  {Yu}}\ and\ \bibinfo {author} {\bibfnamefont {F.}~\bibnamefont {Capasso}},\
  }\href@noop {} {\bibfield  {journal} {\bibinfo  {journal} {Nature materials}\
  }\textbf {\bibinfo {volume} {13}},\ \bibinfo {pages} {139} (\bibinfo {year}
  {2014}{\natexlab{a}})}\BibitemShut {NoStop}%
\bibitem [{\citenamefont {Khorasaninejad}\ \emph
  {et~al.}(2016{\natexlab{a}})\citenamefont {Khorasaninejad}, \citenamefont
  {Chen}, \citenamefont {Devlin}, \citenamefont {Oh}, \citenamefont {Zhu},\
  and\ \citenamefont {Capasso}}]{khorasaninejad2016metalenses}%
  \BibitemOpen
  \bibfield  {author} {\bibinfo {author} {\bibfnamefont {M.}~\bibnamefont
  {Khorasaninejad}}, \bibinfo {author} {\bibfnamefont {W.~T.}\ \bibnamefont
  {Chen}}, \bibinfo {author} {\bibfnamefont {R.~C.}\ \bibnamefont {Devlin}},
  \bibinfo {author} {\bibfnamefont {J.}~\bibnamefont {Oh}}, \bibinfo {author}
  {\bibfnamefont {A.~Y.}\ \bibnamefont {Zhu}}, \ and\ \bibinfo {author}
  {\bibfnamefont {F.}~\bibnamefont {Capasso}},\ }\href@noop {} {\bibfield
  {journal} {\bibinfo  {journal} {Science}\ }\textbf {\bibinfo {volume}
  {352}},\ \bibinfo {pages} {1190} (\bibinfo {year}
  {2016}{\natexlab{a}})}\BibitemShut {NoStop}%
\bibitem [{\citenamefont {Lin}\ \emph {et~al.}(2014)\citenamefont {Lin},
  \citenamefont {Fan}, \citenamefont {Hasman},\ and\ \citenamefont
  {Brongersma}}]{lin2014dielectric}%
  \BibitemOpen
  \bibfield  {author} {\bibinfo {author} {\bibfnamefont {D.}~\bibnamefont
  {Lin}}, \bibinfo {author} {\bibfnamefont {P.}~\bibnamefont {Fan}}, \bibinfo
  {author} {\bibfnamefont {E.}~\bibnamefont {Hasman}}, \ and\ \bibinfo {author}
  {\bibfnamefont {M.~L.}\ \bibnamefont {Brongersma}},\ }\href@noop {}
  {\bibfield  {journal} {\bibinfo  {journal} {Science}\ }\textbf {\bibinfo
  {volume} {345}},\ \bibinfo {pages} {298} (\bibinfo {year}
  {2014})}\BibitemShut {NoStop}%
\bibitem [{\citenamefont {Ni}\ \emph {et~al.}(2012)\citenamefont {Ni},
  \citenamefont {Emani}, \citenamefont {Kildishev}, \citenamefont
  {Boltasseva},\ and\ \citenamefont {Shalaev}}]{ni2012broadband}%
  \BibitemOpen
  \bibfield  {author} {\bibinfo {author} {\bibfnamefont {X.}~\bibnamefont
  {Ni}}, \bibinfo {author} {\bibfnamefont {N.~K.}\ \bibnamefont {Emani}},
  \bibinfo {author} {\bibfnamefont {A.~V.}\ \bibnamefont {Kildishev}}, \bibinfo
  {author} {\bibfnamefont {A.}~\bibnamefont {Boltasseva}}, \ and\ \bibinfo
  {author} {\bibfnamefont {V.~M.}\ \bibnamefont {Shalaev}},\ }\href@noop {}
  {\bibfield  {journal} {\bibinfo  {journal} {Science}\ }\textbf {\bibinfo
  {volume} {335}},\ \bibinfo {pages} {427} (\bibinfo {year}
  {2012})}\BibitemShut {NoStop}%
\bibitem [{\citenamefont {Sun}\ \emph {et~al.}(2012)\citenamefont {Sun},
  \citenamefont {He}, \citenamefont {Xiao}, \citenamefont {Xu}, \citenamefont
  {Li},\ and\ \citenamefont {Zhou}}]{sun2012gradient}%
  \BibitemOpen
  \bibfield  {author} {\bibinfo {author} {\bibfnamefont {S.}~\bibnamefont
  {Sun}}, \bibinfo {author} {\bibfnamefont {Q.}~\bibnamefont {He}}, \bibinfo
  {author} {\bibfnamefont {S.}~\bibnamefont {Xiao}}, \bibinfo {author}
  {\bibfnamefont {Q.}~\bibnamefont {Xu}}, \bibinfo {author} {\bibfnamefont
  {X.}~\bibnamefont {Li}}, \ and\ \bibinfo {author} {\bibfnamefont
  {L.}~\bibnamefont {Zhou}},\ }\href@noop {} {\bibfield  {journal} {\bibinfo
  {journal} {Nature materials}\ }\textbf {\bibinfo {volume} {11}},\ \bibinfo
  {pages} {426} (\bibinfo {year} {2012})}\BibitemShut {NoStop}%
\bibitem [{\citenamefont {Vashistha}\ \emph {et~al.}(2017)\citenamefont
  {Vashistha}, \citenamefont {Vaidya}, \citenamefont {Hegde}, \citenamefont
  {Serebryannikov}, \citenamefont {Bonod},\ and\ \citenamefont
  {Krawczyk}}]{vashistha2017all}%
  \BibitemOpen
  \bibfield  {author} {\bibinfo {author} {\bibfnamefont {V.}~\bibnamefont
  {Vashistha}}, \bibinfo {author} {\bibfnamefont {G.}~\bibnamefont {Vaidya}},
  \bibinfo {author} {\bibfnamefont {R.~S.}\ \bibnamefont {Hegde}}, \bibinfo
  {author} {\bibfnamefont {A.~E.}\ \bibnamefont {Serebryannikov}}, \bibinfo
  {author} {\bibfnamefont {N.}~\bibnamefont {Bonod}}, \ and\ \bibinfo {author}
  {\bibfnamefont {M.}~\bibnamefont {Krawczyk}},\ }\href@noop {} {\bibfield
  {journal} {\bibinfo  {journal} {ACS Photonics}\ }\textbf {\bibinfo {volume}
  {4}},\ \bibinfo {pages} {1076} (\bibinfo {year} {2017})}\BibitemShut
  {NoStop}%
\bibitem [{\citenamefont {Jha}\ \emph {et~al.}(2015)\citenamefont {Jha},
  \citenamefont {Ni}, \citenamefont {Wu}, \citenamefont {Wang},\ and\
  \citenamefont {Zhang}}]{jha2015metasurface}%
  \BibitemOpen
  \bibfield  {author} {\bibinfo {author} {\bibfnamefont {P.~K.}\ \bibnamefont
  {Jha}}, \bibinfo {author} {\bibfnamefont {X.}~\bibnamefont {Ni}}, \bibinfo
  {author} {\bibfnamefont {C.}~\bibnamefont {Wu}}, \bibinfo {author}
  {\bibfnamefont {Y.}~\bibnamefont {Wang}}, \ and\ \bibinfo {author}
  {\bibfnamefont {X.}~\bibnamefont {Zhang}},\ }\href@noop {} {\bibfield
  {journal} {\bibinfo  {journal} {Physical review letters}\ }\textbf {\bibinfo
  {volume} {115}},\ \bibinfo {pages} {025501} (\bibinfo {year}
  {2015})}\BibitemShut {NoStop}%
\bibitem [{\citenamefont {Papp}\ and\ \citenamefont
  {Csaba}(2018)}]{papp2018lens}%
  \BibitemOpen
  \bibfield  {author} {\bibinfo {author} {\bibfnamefont {A.}~\bibnamefont
  {Papp}}\ and\ \bibinfo {author} {\bibfnamefont {G.}~\bibnamefont {Csaba}},\
  }\href@noop {} {\bibfield  {journal} {\bibinfo  {journal} {IEEE Magnetics
  Letters}\ } (\bibinfo {year} {2018})}\BibitemShut {NoStop}%
\bibitem [{\citenamefont {Csaba}\ \emph {et~al.}(2014)\citenamefont {Csaba},
  \citenamefont {Papp},\ and\ \citenamefont {Porod}}]{csaba2014spin}%
  \BibitemOpen
  \bibfield  {author} {\bibinfo {author} {\bibfnamefont {G.}~\bibnamefont
  {Csaba}}, \bibinfo {author} {\bibfnamefont {A.}~\bibnamefont {Papp}}, \ and\
  \bibinfo {author} {\bibfnamefont {W.}~\bibnamefont {Porod}},\ }\href@noop {}
  {\bibfield  {journal} {\bibinfo  {journal} {Journal of Applied Physics}\
  }\textbf {\bibinfo {volume} {115}},\ \bibinfo {pages} {17C741} (\bibinfo
  {year} {2014})}\BibitemShut {NoStop}%
\bibitem [{\citenamefont {Toedt}\ \emph {et~al.}(2016)\citenamefont {Toedt},
  \citenamefont {Mundkowski}, \citenamefont {Heitmann}, \citenamefont
  {Mendach},\ and\ \citenamefont {Hansen}}]{toedt2016design}%
  \BibitemOpen
  \bibfield  {author} {\bibinfo {author} {\bibfnamefont {J.-N.}\ \bibnamefont
  {Toedt}}, \bibinfo {author} {\bibfnamefont {M.}~\bibnamefont {Mundkowski}},
  \bibinfo {author} {\bibfnamefont {D.}~\bibnamefont {Heitmann}}, \bibinfo
  {author} {\bibfnamefont {S.}~\bibnamefont {Mendach}}, \ and\ \bibinfo
  {author} {\bibfnamefont {W.}~\bibnamefont {Hansen}},\ }\href@noop {}
  {\bibfield  {journal} {\bibinfo  {journal} {Scientific reports}\ }\textbf
  {\bibinfo {volume} {6}},\ \bibinfo {pages} {33169} (\bibinfo {year}
  {2016})}\BibitemShut {NoStop}%
\bibitem [{\citenamefont {Whitehead}\ \emph {et~al.}(2018)\citenamefont
  {Whitehead}, \citenamefont {Horsley}, \citenamefont {Philbin},\ and\
  \citenamefont {Kruglyak}}]{whitehead2018luneburg}%
  \BibitemOpen
  \bibfield  {author} {\bibinfo {author} {\bibfnamefont {N.~J.}\ \bibnamefont
  {Whitehead}}, \bibinfo {author} {\bibfnamefont {S.~A.~R.}\ \bibnamefont
  {Horsley}}, \bibinfo {author} {\bibfnamefont {T.~G.}\ \bibnamefont
  {Philbin}}, \ and\ \bibinfo {author} {\bibfnamefont {V.~V.}\ \bibnamefont
  {Kruglyak}},\ }\href@noop {} {\bibfield  {journal} {\bibinfo  {journal}
  {Applied Physics Letters}\ }\textbf {\bibinfo {volume} {113}},\ \bibinfo
  {pages} {212404} (\bibinfo {year} {2018})}\BibitemShut {NoStop}%
\bibitem [{\citenamefont {Dzyapko}\ \emph {et~al.}(2016)\citenamefont
  {Dzyapko}, \citenamefont {Borisenko}, \citenamefont {Demidov}, \citenamefont
  {Pernice},\ and\ \citenamefont {Demokritov}}]{dzyapko2016reconfigurable}%
  \BibitemOpen
  \bibfield  {author} {\bibinfo {author} {\bibfnamefont {O.}~\bibnamefont
  {Dzyapko}}, \bibinfo {author} {\bibfnamefont {I.}~\bibnamefont {Borisenko}},
  \bibinfo {author} {\bibfnamefont {V.}~\bibnamefont {Demidov}}, \bibinfo
  {author} {\bibfnamefont {W.}~\bibnamefont {Pernice}}, \ and\ \bibinfo
  {author} {\bibfnamefont {S.}~\bibnamefont {Demokritov}},\ }\href@noop {}
  {\bibfield  {journal} {\bibinfo  {journal} {Applied Physics Letters}\
  }\textbf {\bibinfo {volume} {109}},\ \bibinfo {pages} {232407} (\bibinfo
  {year} {2016})}\BibitemShut {NoStop}%
\bibitem [{\citenamefont {Gruszecki}\ and\ \citenamefont
  {Krawczyk}(2018)}]{gruszecki2018mirage}%
  \BibitemOpen
  \bibfield  {author} {\bibinfo {author} {\bibfnamefont {P.}~\bibnamefont
  {Gruszecki}}\ and\ \bibinfo {author} {\bibfnamefont {M.}~\bibnamefont
  {Krawczyk}},\ }\href@noop {} {\bibfield  {journal} {\bibinfo  {journal}
  {Physical Review B}\ }\textbf {\bibinfo {volume} {97}},\ \bibinfo {pages}
  {094424} (\bibinfo {year} {2018})}\BibitemShut {NoStop}%
\bibitem [{\citenamefont {Bruno}(1993)}]{Bruno1993InterlayerPicture}%
  \BibitemOpen
  \bibfield  {author} {\bibinfo {author} {\bibfnamefont {P.}~\bibnamefont
  {Bruno}},\ }\href {\doibase 10.1016/0304-8853(93)91197-F} {\bibfield
  {journal} {\bibinfo  {journal} {Journal of Magnetism and Magnetic Materials}\
  }\textbf {\bibinfo {volume} {121}},\ \bibinfo {pages} {248} (\bibinfo {year}
  {1993})}\BibitemShut {NoStop}%
\bibitem [{\citenamefont {Bruno}(1995)}]{Bruno1995QuantumCo/Cu/Co001}%
  \BibitemOpen
  \bibfield  {author} {\bibinfo {author} {\bibfnamefont {P.}~\bibnamefont
  {Bruno}},\ }\href {\doibase 10.1016/0304-8853(95)00208-1} {\bibfield
  {journal} {\bibinfo  {journal} {Journal of Magnetism and Magnetic Materials}\
  }\textbf {\bibinfo {volume} {148}},\ \bibinfo {pages} {202} (\bibinfo {year}
  {1995})}\BibitemShut {NoStop}%
\bibitem [{\citenamefont {Bruno}\ and\ \citenamefont
  {Chappert}(1991)}]{Bruno1991OscillatorySpacer}%
  \BibitemOpen
  \bibfield  {author} {\bibinfo {author} {\bibfnamefont {P.}~\bibnamefont
  {Bruno}}\ and\ \bibinfo {author} {\bibfnamefont {C.}~\bibnamefont
  {Chappert}},\ }\href {\doibase 10.1103/PhysRevLett.67.1602} {\bibfield
  {journal} {\bibinfo  {journal} {Physical Review Letters}\ }\textbf {\bibinfo
  {volume} {67}},\ \bibinfo {pages} {1602} (\bibinfo {year}
  {1991})}\BibitemShut {NoStop}%
\bibitem [{\citenamefont {Ruderman}\ and\ \citenamefont
  {Kittel}(1954)}]{Ruderman1954IndirectElectrons}%
  \BibitemOpen
  \bibfield  {author} {\bibinfo {author} {\bibfnamefont {M.~A.}\ \bibnamefont
  {Ruderman}}\ and\ \bibinfo {author} {\bibfnamefont {C.}~\bibnamefont
  {Kittel}},\ }\href {\doibase 10.1103/PhysRev.96.99} {\bibfield  {journal}
  {\bibinfo  {journal} {Physical Review}\ }\textbf {\bibinfo {volume} {96}},\
  \bibinfo {pages} {99} (\bibinfo {year} {1954})}\BibitemShut {NoStop}%
\bibitem [{\citenamefont {Chappert}\ and\ \citenamefont
  {Renard}(1991)}]{Chappert1991Long-periodPicture}%
  \BibitemOpen
  \bibfield  {author} {\bibinfo {author} {\bibfnamefont {C.}~\bibnamefont
  {Chappert}}\ and\ \bibinfo {author} {\bibfnamefont {J.~P.}\ \bibnamefont
  {Renard}},\ }\href {\doibase 10.1209/0295-5075/15/5/014} {\bibfield
  {journal} {\bibinfo  {journal} {Epl}\ }\textbf {\bibinfo {volume} {15}},\
  \bibinfo {pages} {553} (\bibinfo {year} {1991})}\BibitemShut {NoStop}%
\bibitem [{\citenamefont {Bruno}\ and\ \citenamefont
  {Chappert}(1992)}]{Bruno1992Ruderman-KittelCoupling}%
  \BibitemOpen
  \bibfield  {author} {\bibinfo {author} {\bibfnamefont {P.}~\bibnamefont
  {Bruno}}\ and\ \bibinfo {author} {\bibfnamefont {C.}~\bibnamefont
  {Chappert}},\ }\href {\doibase 10.1103/PhysRevB.46.261} {\bibfield  {journal}
  {\bibinfo  {journal} {Physical Review B}\ }\textbf {\bibinfo {volume} {46}},\
  \bibinfo {pages} {261} (\bibinfo {year} {1992})}\BibitemShut {NoStop}%
\bibitem [{\citenamefont {Johnson}\ \emph {et~al.}(1992)\citenamefont
  {Johnson}, \citenamefont {Purcell}, \citenamefont {McGee}, \citenamefont
  {Coehoorn}, \citenamefont {aan~de Stegge},\ and\ \citenamefont
  {Hoving}}]{Johnson1992StructuralLayers}%
  \BibitemOpen
  \bibfield  {author} {\bibinfo {author} {\bibfnamefont {M.~T.}\ \bibnamefont
  {Johnson}}, \bibinfo {author} {\bibfnamefont {S.~T.}\ \bibnamefont
  {Purcell}}, \bibinfo {author} {\bibfnamefont {N.~W.~E.}\ \bibnamefont
  {McGee}}, \bibinfo {author} {\bibfnamefont {R.}~\bibnamefont {Coehoorn}},
  \bibinfo {author} {\bibfnamefont {J.}~\bibnamefont {aan~de Stegge}}, \ and\
  \bibinfo {author} {\bibfnamefont {W.}~\bibnamefont {Hoving}},\ }\href
  {\doibase 10.1103/PhysRevLett.68.2688} {\bibfield  {journal} {\bibinfo
  {journal} {Physical Review Letters}\ }\textbf {\bibinfo {volume} {68}},\
  \bibinfo {pages} {2688} (\bibinfo {year} {1992})}\BibitemShut {NoStop}%
\bibitem [{\citenamefont {Bloemen}\ \emph {et~al.}(1993)\citenamefont
  {Bloemen}, \citenamefont {van Dalen}, \citenamefont {de~Jonge}, \citenamefont
  {Johnson},\ and\ \citenamefont {aan~de
  Stegge}}]{Bloemen1993ShortCo/Cu/Co100}%
  \BibitemOpen
  \bibfield  {author} {\bibinfo {author} {\bibfnamefont {P.~J.~H.}\
  \bibnamefont {Bloemen}}, \bibinfo {author} {\bibfnamefont {R.}~\bibnamefont
  {van Dalen}}, \bibinfo {author} {\bibfnamefont {W.~J.~M.}\ \bibnamefont
  {de~Jonge}}, \bibinfo {author} {\bibfnamefont {M.~T.}\ \bibnamefont
  {Johnson}}, \ and\ \bibinfo {author} {\bibfnamefont {J.}~\bibnamefont {aan~de
  Stegge}},\ }\href {\doibase 10.1063/1.353487} {\bibfield  {journal} {\bibinfo
   {journal} {Journal of Applied Physics}\ }\textbf {\bibinfo {volume} {73}},\
  \bibinfo {pages} {5972} (\bibinfo {year} {1993})}\BibitemShut {NoStop}%
\bibitem [{\citenamefont {Kruglyak}\ \emph {et~al.}(2014)\citenamefont
  {Kruglyak}, \citenamefont {Gorobets}, \citenamefont {Gorobets},\ and\
  \citenamefont {Kuchko}}]{Kruglyak2014MagnetizationThickness}%
  \BibitemOpen
  \bibfield  {author} {\bibinfo {author} {\bibfnamefont {V.~V.}\ \bibnamefont
  {Kruglyak}}, \bibinfo {author} {\bibfnamefont {O.~Y.}\ \bibnamefont
  {Gorobets}}, \bibinfo {author} {\bibfnamefont {Y.~I.}\ \bibnamefont
  {Gorobets}}, \ and\ \bibinfo {author} {\bibfnamefont {A.~N.}\ \bibnamefont
  {Kuchko}},\ }\href {\doibase 10.1088/0953-8984/26/40/406001} {\bibfield
  {journal} {\bibinfo  {journal} {Journal of Physics Condensed Matter}\
  }\textbf {\bibinfo {volume} {26}} (\bibinfo {year} {2014}),\
  10.1088/0953-8984/26/40/406001}\BibitemShut {NoStop}%
\bibitem [{\citenamefont {Gruszecki}\ \emph {et~al.}(2017)\citenamefont
  {Gruszecki}, \citenamefont {Mailyan}, \citenamefont {Gorobets},\ and\
  \citenamefont {Krawczyk}}]{gruszecki2017goos}%
  \BibitemOpen
  \bibfield  {author} {\bibinfo {author} {\bibfnamefont {P.}~\bibnamefont
  {Gruszecki}}, \bibinfo {author} {\bibfnamefont {M.}~\bibnamefont {Mailyan}},
  \bibinfo {author} {\bibfnamefont {O.}~\bibnamefont {Gorobets}}, \ and\
  \bibinfo {author} {\bibfnamefont {M.}~\bibnamefont {Krawczyk}},\ }\href@noop
  {} {\bibfield  {journal} {\bibinfo  {journal} {Physical Review B}\ }\textbf
  {\bibinfo {volume} {95}},\ \bibinfo {pages} {014421} (\bibinfo {year}
  {2017})}\BibitemShut {NoStop}%
\bibitem [{\citenamefont {Vansteenkiste}\ and\ \citenamefont {Van~de
  Wiele}(2011)}]{MuMax2011_main}%
  \BibitemOpen
  \bibfield  {author} {\bibinfo {author} {\bibfnamefont {A.}~\bibnamefont
  {Vansteenkiste}}\ and\ \bibinfo {author} {\bibfnamefont {B.}~\bibnamefont
  {Van~de Wiele}},\ }\href {\doibase 10.1016/j.jmmm.2011.05.037} {\bibfield
  {journal} {\bibinfo  {journal} {Journal of Magnetism and Magnetic Materials}\
  }\textbf {\bibinfo {volume} {323}},\ \bibinfo {pages} {2585} (\bibinfo {year}
  {2011})}\BibitemShut {NoStop}%
\bibitem [{\citenamefont {Vansteenkiste}\ \emph {et~al.}(2014)\citenamefont
  {Vansteenkiste}, \citenamefont {Leliaert}, \citenamefont {Dvornik},
  \citenamefont {Helsen}, \citenamefont {Garcia-Sanchez},\ and\ \citenamefont
  {Van~Waeyenberge}}]{mumax_2014}%
  \BibitemOpen
  \bibfield  {author} {\bibinfo {author} {\bibfnamefont {A.}~\bibnamefont
  {Vansteenkiste}}, \bibinfo {author} {\bibfnamefont {J.}~\bibnamefont
  {Leliaert}}, \bibinfo {author} {\bibfnamefont {M.}~\bibnamefont {Dvornik}},
  \bibinfo {author} {\bibfnamefont {M.}~\bibnamefont {Helsen}}, \bibinfo
  {author} {\bibfnamefont {F.}~\bibnamefont {Garcia-Sanchez}}, \ and\ \bibinfo
  {author} {\bibfnamefont {B.}~\bibnamefont {Van~Waeyenberge}},\ }\href
  {\doibase 10.1063/1.4899186} {\bibfield  {journal} {\bibinfo  {journal} {AIP
  Advances}\ }\textbf {\bibinfo {volume} {4}},\ \bibinfo {pages} {107133}
  (\bibinfo {year} {2014})}\BibitemShut {NoStop}%
\bibitem [{\citenamefont {Mulkers}\ \emph {et~al.}(2017)\citenamefont
  {Mulkers}, \citenamefont {Van~Waeyenberge},\ and\ \citenamefont
  {Milo{\v{s}}evi{\'{c}}}}]{Mulkers2017}%
  \BibitemOpen
  \bibfield  {author} {\bibinfo {author} {\bibfnamefont {J.}~\bibnamefont
  {Mulkers}}, \bibinfo {author} {\bibfnamefont {B.}~\bibnamefont
  {Van~Waeyenberge}}, \ and\ \bibinfo {author} {\bibfnamefont {M.~V.}\
  \bibnamefont {Milo{\v{s}}evi{\'{c}}}},\ }\href {\doibase
  10.1103/PhysRevB.95.144401} {\bibfield  {journal} {\bibinfo  {journal}
  {Physical Review B}\ }\textbf {\bibinfo {volume} {95}},\ \bibinfo {pages}
  {144401} (\bibinfo {year} {2017})}\BibitemShut {NoStop}%
\bibitem [{\citenamefont {Exl}\ \emph {et~al.}(2014)\citenamefont {Exl},
  \citenamefont {Bance}, \citenamefont {Reichel}, \citenamefont {Schrefl},
  \citenamefont {Peter~Stimming},\ and\ \citenamefont {Mauser}}]{Exl2014}%
  \BibitemOpen
  \bibfield  {author} {\bibinfo {author} {\bibfnamefont {L.}~\bibnamefont
  {Exl}}, \bibinfo {author} {\bibfnamefont {S.}~\bibnamefont {Bance}}, \bibinfo
  {author} {\bibfnamefont {F.}~\bibnamefont {Reichel}}, \bibinfo {author}
  {\bibfnamefont {T.}~\bibnamefont {Schrefl}}, \bibinfo {author} {\bibfnamefont
  {H.}~\bibnamefont {Peter~Stimming}}, \ and\ \bibinfo {author} {\bibfnamefont
  {N.~J.}\ \bibnamefont {Mauser}},\ }\href {\doibase 10.1063/1.4862839}
  {\bibfield  {journal} {\bibinfo  {journal} {Journal of Applied Physics}\
  }\textbf {\bibinfo {volume} {115}},\ \bibinfo {pages} {17D118} (\bibinfo
  {year} {2014})}\BibitemShut {NoStop}%
\bibitem [{\citenamefont {Leliaert}\ \emph {et~al.}(2014)\citenamefont
  {Leliaert}, \citenamefont {Van~de Wiele}, \citenamefont {Vansteenkiste},
  \citenamefont {Laurson}, \citenamefont {Durin}, \citenamefont {Dupr{\'{e}}},\
  and\ \citenamefont {Van~Waeyenberge}}]{Leliaert2014}%
  \BibitemOpen
  \bibfield  {author} {\bibinfo {author} {\bibfnamefont {J.}~\bibnamefont
  {Leliaert}}, \bibinfo {author} {\bibfnamefont {B.}~\bibnamefont {Van~de
  Wiele}}, \bibinfo {author} {\bibfnamefont {A.}~\bibnamefont {Vansteenkiste}},
  \bibinfo {author} {\bibfnamefont {L.}~\bibnamefont {Laurson}}, \bibinfo
  {author} {\bibfnamefont {G.}~\bibnamefont {Durin}}, \bibinfo {author}
  {\bibfnamefont {L.}~\bibnamefont {Dupr{\'{e}}}}, \ and\ \bibinfo {author}
  {\bibfnamefont {B.}~\bibnamefont {Van~Waeyenberge}},\ }\href@noop {}
  {\bibfield  {journal} {\bibinfo  {journal} {Journal of Applied Physics}\
  }\textbf {\bibinfo {volume} {115}} (\bibinfo {year} {2014})}\BibitemShut
  {NoStop}%
\bibitem [{\citenamefont {Venkat}\ \emph {et~al.}(2018)\citenamefont {Venkat},
  \citenamefont {Fangohr},\ and\ \citenamefont
  {Prabhakar}}]{venkat2018absorbing}%
  \BibitemOpen
  \bibfield  {author} {\bibinfo {author} {\bibfnamefont {G.}~\bibnamefont
  {Venkat}}, \bibinfo {author} {\bibfnamefont {H.}~\bibnamefont {Fangohr}}, \
  and\ \bibinfo {author} {\bibfnamefont {A.}~\bibnamefont {Prabhakar}},\
  }\href@noop {} {\bibfield  {journal} {\bibinfo  {journal} {Journal of
  Magnetism and Magnetic Materials}\ }\textbf {\bibinfo {volume} {450}},\
  \bibinfo {pages} {34} (\bibinfo {year} {2018})}\BibitemShut {NoStop}%
\bibitem [{Note1()}]{Note1}%
  \BibitemOpen
  \bibinfo {note} {We used a uniformly discretizing grid with the cell size
  0.5-1.0 nm $\times $ 0.5-1.0 nm $\times $ 10 nm along the $x$, $y$ and $z$
  axis, respectively.}\BibitemShut {Stop}%
\bibitem [{\citenamefont {Chen}\ and\ \citenamefont
  {Han}(2010)}]{Chen2010MicrowaveAnisotropy}%
  \BibitemOpen
  \bibfield  {author} {\bibinfo {author} {\bibfnamefont {W.}~\bibnamefont
  {Chen}}\ and\ \bibinfo {author} {\bibfnamefont {M.}~\bibnamefont {Han}},\
  }\href {\doibase 10.2529/PIERS090824024047} {\bibfield  {journal} {\bibinfo
  {journal} {PIERS Online}\ }\textbf {\bibinfo {volume} {6}},\ \bibinfo {pages}
  {101} (\bibinfo {year} {2010})}\BibitemShut {NoStop}%
\bibitem [{\citenamefont {Cullity}\ and\ \citenamefont
  {Graham}(2008)}]{Cullity2008IntroductionMaterials}%
  \BibitemOpen
  \bibfield  {author} {\bibinfo {author} {\bibfnamefont {B.~D.}\ \bibnamefont
  {Cullity}}\ and\ \bibinfo {author} {\bibfnamefont {C.~D.}\ \bibnamefont
  {Graham}},\ }\href {\doibase 10.1002/9780470386323} {\emph {\bibinfo {title}
  {Introduction to Magnetic Materials}}},\ \bibinfo {edition} {2nd}\ ed.\
  (\bibinfo  {publisher} {Wiley-IEEE Press},\ \bibinfo {year} {2008})\ p.\
  \bibinfo {pages} {221}\BibitemShut {NoStop}%
\bibitem [{Note2()}]{Note2}%
  \BibitemOpen
  \bibinfo {note} {The recent investigations Ref.\cite
  {Szalowski2008PRB_RKKY,Hjorvarsson1997PRL_Reversible,Klose1997PRL_Continous,Unguris1997PRL_Determination}
  show that, the smooth changes of the RKKY interactions are
  feasible.}\BibitemShut {Stop}%
\bibitem [{\citenamefont {Stigloher}\ \emph {et~al.}(2016)\citenamefont
  {Stigloher}, \citenamefont {Decker}, \citenamefont {K{\"o}rner},
  \citenamefont {Tanabe}, \citenamefont {Moriyama}, \citenamefont {Taniguchi},
  \citenamefont {Hata}, \citenamefont {Madami}, \citenamefont {Gubbiotti},
  \citenamefont {Kobayashi} \emph {et~al.}}]{stigloher2016snell}%
  \BibitemOpen
  \bibfield  {author} {\bibinfo {author} {\bibfnamefont {J.}~\bibnamefont
  {Stigloher}}, \bibinfo {author} {\bibfnamefont {M.}~\bibnamefont {Decker}},
  \bibinfo {author} {\bibfnamefont {H.~S.}\ \bibnamefont {K{\"o}rner}},
  \bibinfo {author} {\bibfnamefont {K.}~\bibnamefont {Tanabe}}, \bibinfo
  {author} {\bibfnamefont {T.}~\bibnamefont {Moriyama}}, \bibinfo {author}
  {\bibfnamefont {T.}~\bibnamefont {Taniguchi}}, \bibinfo {author}
  {\bibfnamefont {H.}~\bibnamefont {Hata}}, \bibinfo {author} {\bibfnamefont
  {M.}~\bibnamefont {Madami}}, \bibinfo {author} {\bibfnamefont
  {G.}~\bibnamefont {Gubbiotti}}, \bibinfo {author} {\bibfnamefont
  {K.}~\bibnamefont {Kobayashi}},  \emph {et~al.},\ }\href@noop {} {\bibfield
  {journal} {\bibinfo  {journal} {Physical review letters}\ }\textbf {\bibinfo
  {volume} {117}},\ \bibinfo {pages} {037204} (\bibinfo {year}
  {2016})}\BibitemShut {NoStop}%
\bibitem [{\citenamefont {Yu}\ and\ \citenamefont
  {Capasso}(2014{\natexlab{b}})}]{Yu2014}%
  \BibitemOpen
  \bibfield  {author} {\bibinfo {author} {\bibfnamefont {N.}~\bibnamefont
  {Yu}}\ and\ \bibinfo {author} {\bibfnamefont {F.}~\bibnamefont {Capasso}},\
  }\href {\doibase 10.1038/nmat3839} {\bibfield  {journal} {\bibinfo  {journal}
  {Nature Materials}\ }\textbf {\bibinfo {volume} {13}},\ \bibinfo {pages}
  {139} (\bibinfo {year} {2014}{\natexlab{b}})}\BibitemShut {NoStop}%
\bibitem [{Note3()}]{Note3}%
  \BibitemOpen
  \bibinfo {note} {We have used the modified code of the mumax3 allowing us to
  use more regions to implement the scale exchange interface. The limit in the
  standard version of mumax3 is 256 regions.}\BibitemShut {Stop}%
\bibitem [{\citenamefont {Khorasaninejad}\ \emph
  {et~al.}(2016{\natexlab{b}})\citenamefont {Khorasaninejad}, \citenamefont
  {Zhu}, \citenamefont {Roques-Carmes}, \citenamefont {Chen}, \citenamefont
  {Oh}, \citenamefont {Mishra}, \citenamefont {Devlin},\ and\ \citenamefont
  {Capasso}}]{Khorasaninejad2016Polarization-InsensitiveWavelengths}%
  \BibitemOpen
  \bibfield  {author} {\bibinfo {author} {\bibfnamefont {M.}~\bibnamefont
  {Khorasaninejad}}, \bibinfo {author} {\bibfnamefont {A.~Y.}\ \bibnamefont
  {Zhu}}, \bibinfo {author} {\bibfnamefont {C.}~\bibnamefont {Roques-Carmes}},
  \bibinfo {author} {\bibfnamefont {W.~T.}\ \bibnamefont {Chen}}, \bibinfo
  {author} {\bibfnamefont {J.}~\bibnamefont {Oh}}, \bibinfo {author}
  {\bibfnamefont {I.}~\bibnamefont {Mishra}}, \bibinfo {author} {\bibfnamefont
  {R.~C.}\ \bibnamefont {Devlin}}, \ and\ \bibinfo {author} {\bibfnamefont
  {F.}~\bibnamefont {Capasso}},\ }\href {\doibase 10.1021/acs.nanolett.6b03626}
  {\bibfield  {journal} {\bibinfo  {journal} {Nano Letters}\ }\textbf {\bibinfo
  {volume} {16}},\ \bibinfo {pages} {7229} (\bibinfo {year}
  {2016}{\natexlab{b}})}\BibitemShut {NoStop}%
\bibitem [{\citenamefont {Sza{\l}owski}\ and\ \citenamefont
  {Balcerzak}(2008)}]{Szalowski2008PRB_RKKY}%
  \BibitemOpen
  \bibfield  {author} {\bibinfo {author} {\bibfnamefont {K.}~\bibnamefont
  {Sza{\l}owski}}\ and\ \bibinfo {author} {\bibfnamefont {T.}~\bibnamefont
  {Balcerzak}},\ }\href {\doibase 10.1103/PhysRevB.78.024419} {\bibfield
  {journal} {\bibinfo  {journal} {Physical Review B - Condensed Matter and
  Materials Physics}\ }\textbf {\bibinfo {volume} {78}},\ \bibinfo {pages} {1}
  (\bibinfo {year} {2008})}\BibitemShut {NoStop}%
\bibitem [{\citenamefont {Hj{\"{o}}rvarsson}\ \emph {et~al.}(1997)\citenamefont
  {Hj{\"{o}}rvarsson}, \citenamefont {Dura}, \citenamefont {Isberg},
  \citenamefont {Watanabe}, \citenamefont {Udovic}, \citenamefont {Andersson},\
  and\ \citenamefont {Majkrzak}}]{Hjorvarsson1997PRL_Reversible}%
  \BibitemOpen
  \bibfield  {author} {\bibinfo {author} {\bibfnamefont {B.}~\bibnamefont
  {Hj{\"{o}}rvarsson}}, \bibinfo {author} {\bibfnamefont {J.~A.}\ \bibnamefont
  {Dura}}, \bibinfo {author} {\bibfnamefont {P.}~\bibnamefont {Isberg}},
  \bibinfo {author} {\bibfnamefont {T.}~\bibnamefont {Watanabe}}, \bibinfo
  {author} {\bibfnamefont {T.~J.}\ \bibnamefont {Udovic}}, \bibinfo {author}
  {\bibfnamefont {G.}~\bibnamefont {Andersson}}, \ and\ \bibinfo {author}
  {\bibfnamefont {C.~F.}\ \bibnamefont {Majkrzak}},\ }\href {\doibase
  10.1103/PhysRevLett.79.901} {\bibfield  {journal} {\bibinfo  {journal}
  {Physical Review Letters}\ }\textbf {\bibinfo {volume} {79}},\ \bibinfo
  {pages} {901} (\bibinfo {year} {1997})}\BibitemShut {NoStop}%
\bibitem [{\citenamefont {Klose}\ \emph {et~al.}(1997)\citenamefont {Klose},
  \citenamefont {Rehm}, \citenamefont {Nagengast}, \citenamefont {Maletta},\
  and\ \citenamefont {Weidinger}}]{Klose1997PRL_Continous}%
  \BibitemOpen
  \bibfield  {author} {\bibinfo {author} {\bibfnamefont {F.}~\bibnamefont
  {Klose}}, \bibinfo {author} {\bibfnamefont {C.}~\bibnamefont {Rehm}},
  \bibinfo {author} {\bibfnamefont {D.}~\bibnamefont {Nagengast}}, \bibinfo
  {author} {\bibfnamefont {H.}~\bibnamefont {Maletta}}, \ and\ \bibinfo
  {author} {\bibfnamefont {A.}~\bibnamefont {Weidinger}},\ }\href {\doibase
  10.1103/PhysRevLett.78.1150} {\bibfield  {journal} {\bibinfo  {journal}
  {Physical Review Letters}\ }\textbf {\bibinfo {volume} {78}},\ \bibinfo
  {pages} {1150} (\bibinfo {year} {1997})}\BibitemShut {NoStop}%
\bibitem [{\citenamefont {Unguris}\ \emph {et~al.}(1997)\citenamefont
  {Unguris}, \citenamefont {Celotta},\ and\ \citenamefont
  {Pierce}}]{Unguris1997PRL_Determination}%
  \BibitemOpen
  \bibfield  {author} {\bibinfo {author} {\bibfnamefont {J.}~\bibnamefont
  {Unguris}}, \bibinfo {author} {\bibfnamefont {R.~J.}\ \bibnamefont
  {Celotta}}, \ and\ \bibinfo {author} {\bibfnamefont {D.~T.}\ \bibnamefont
  {Pierce}},\ }\href@noop {} {\bibfield  {journal} {\bibinfo  {journal}
  {Physical Review Letters}\ }\textbf {\bibinfo {volume} {1}},\ \bibinfo
  {pages} {1} (\bibinfo {year} {1997})}\BibitemShut {NoStop}%
\end{thebibliography}%

\end{document}